\documentclass[12pt]{article}
\usepackage{amsmath}
\usepackage{amssymb}
\usepackage{amstext}
\usepackage{amsfonts}
\usepackage{graphicx,epsfig}
\usepackage{slashbox}
\usepackage{indentfirst}

\begin{document}

\baselineskip 20pt

\title{Determining the nuclear form factor for detection of dark matter in the relativistic mean field theory }
\author{Ya-Zheng Chen\footnote{cyzufo@mail.nankai.edu.cn}\ ,
        Yan-An Luo\footnote{luoya@nankai.edu.cn}\ ,
        Lei Li\footnote{lilei@nankai.edu.cn}\ ,
        Hong Shen \footnote{songtc@nankai.edu.cn} and
        Xue-Qian Li \footnote{lixq@nankai.edu.cn}\\
{\small  School of Physics, Nankai University, 300071, Tianjin,
China} }
\date{\today}
\maketitle

\begin{abstract}

In this work, we derive the nuclear form factor for the
spin-independent collision between the WIMPs and nucleus in terms of
the relativistic mean field (RMF) theory. Comparison  with the
traditional form factors which are commonly used in literature is
given and it is found that our results are slightly above that of
the 2PF model by 4\% to 8\%, but deviate from the Helm form factor
by 15\% to 25\% for the whole recoil energy spectrum of 0$\sim $ 100
keV. Moreover, taking Xe and Ge as examples, we show the dependence
of the form factor on the recoil energy.

\end{abstract}
\section{Introduction}

Discovery of dark matter undoubtedly was one of the greatest
scientific events of the 20th century, then directly searching for
dark matter and identifying it will be the most important task of
this century. Commonly accepted point of view~\cite{V.Trimble} is
that the main fraction of the dark matter in our universe which
resides at the hallo of the galaxy is the cold dark matter, i.e. the
weakly-interacting-massive-particles (WIMPs), and the most favorable
candidate of the WIMPs is the neutralino~\cite{EM}, even though one
cannot exclude other possibilities at the present~\cite{Lars
Bergstrom,J.L.Feng}. What we expect is to \textquotedblleft
see\textquotedblright\ the dark matter particles in direct search
experiments and find them at the
LHC~\cite{Gaitskell:2004gd,Baer:2004qq,Baer:2009bu}. The important
step is to set experiments to directly search for such mysterious
particles. Since the dark matter particles only participate in weak
and gravitational interactions, the scattering cross section with
normal matter is very small and makes the observation very
difficult. Moreover, even though the dark matter particles are very
heavy, its dispersive velocity is only $200\sim 600$
km/s~\cite{velocity}, thus the available kinetic energy $M(v/c)^2/2$
is of order of keV. This kinetic energy is not enough to make any
inelastic nuclear reaction such as level transitions, therefore the
only effects which are observable are the phenomena related to the
nucleus recoil. This makes the measurement even more difficult in
comparison with the neutrino experiments.

The observation rates for spin-independent scattering can be written
as~\cite{Jungman:1995}
\begin{eqnarray}\label{event}
\frac{d\sigma}{d|{\bf{p}}|^{2}}&=&G_{F}^{2}\frac{C}{v^2}F^{2}(|{\bf{q}}|)
=\frac{\sigma_{0}}{4m_{r}^{2}v^{2}}F^{2}(|{\bf{q}}|),
\end{eqnarray}
where $G_F$ is the universal Fermi coupling constant, $\sigma_0$ is the cross section at zero-recoil, $m_r$
is the reduced mass of the WIMP and nucleus, finally $F^{2}(|{\bf{q}}|)$ is the nuclear form factor.
Generally speaking, the mass density distribution of the nucleus is proportional to the charge
density or the nucleon number density, hence the form factor can also be accounted from the nucleon number density
in the nucleus, i.e. the nuclear density.
The form factor is the Fourier
transformation of the nuclear density as
\begin{eqnarray}\label{fourier transform}
F(q)&=&\frac{1}{A}\int \rho(r)e^{-i{\bf{q}}\cdot
{\bf{r}}}d^3r \nonumber\\
&=&\frac{4\pi}{A}\int \frac{r}{q} \rho(r)\sin{(qr)}dr,
\end{eqnarray}
where $\rho(r)$ is the nuclear density. Here we assume that the nucleus is spherically symmetric and it is only a function of $r$.

There are several ansatz for determining the form factor by assuming
typical $\rho(r)$ functions~\cite{Helm,SOGmodel,FBmodel,JDL}. We will
give a brief review of the more commonly adopted form factors in
next section.

The importance of the form factor is obvious. In fact, one of the
the goals of particle-cosmology is to identify the dark matter. The
strategy is to theoretically calculate $\sigma_0$ in terms of
various models for the WIMPs, which indeed determine how WIMPs
interact with quarks in nucleon. The concrete models possess several
model parameters~\cite{Goodman:1984dc,Drees:1993bu}. Then one can
compare the theoretical results with the data to extract information
about the interaction and identity of the dark matter particle which
is included in $\sigma_0$. However, in Eq. (\ref{event}), $\sigma_0$
is convoluted with the nuclear form factor $F(q)$ which is required
to obey
$$F(0)=1.$$ Unless one can more accurately determine $F(q)$, extraction of
useful information from data is impossible. Thus it is crucially important to determine the form factor. Since the recoil energy spectrum may span an energy range from a few hundreds
eV to a few hundreds keV and the shape of the form factor may have different response to different recoil energies, so it is extremely important to have an accurate $F(q)$ for the full energy range.
Namely, an accurate determination of not only the values, but also the shape of the $F(q)$ is necessary.

In this work, we employ the relativistic mean field (RMF) theory to
determine the nuclear form factors for Xe, Ge and several other
nuclei which can be used as the  materials for the dark matter
detectors. Thus one can first choose any potential as the initial
input, by iteration he eventually finds a stable form which makes
all relevant equations to be self-consistent (see the text for
details), the resultant form does not depend on the initial input
which can be pre-assumed by our knowledge and intuition of physics.
No doubt, if the initial potential is properly chosen, the iteration
would converge quickly. Generally such a resultant form factor
cannot be analytical, but only be presented numerically. With
powerful computers, its application for extracting information of
$\sigma_0$ from data  is not affected at all.

The article is organized as follows, after this introduction, we
briefly review the forms of the nuclear form factors which are
commonly adopted in literature, and then we derive necessary
formulae  for obtaining the form factor in terms of the RMF theory.
In later section, we will present our results and make a comparison
with the Helm and 2PF form factors in several figures. The last
section is devoted to the discussion and analysis.

\section{Nuclear form factor for the spin-independent coupling}

The nuclear form factor is the Fourier transformation of the nuclear
density. In this paper, we numerically calculate the nuclear
density by means of the RMF theory. Namely, we are going to derive the nucleon number density (nuclear density)
in Eq. (\ref{fourier transform}) in terms of the RMF theory.

\subsection{Several types of nuclear form factors}
Here let us briefly  review the form factors for the spin
independent WIMP-nucleus interaction which were commonly adopted in
literature.

1. The simplest form factor is the exponential
form~\cite{Ahlen,Freese}:
\begin{equation}
F(E)=\exp(-E/2Q_0),
\end{equation}
where $E=q^2/2M_A$ is the transferred energy from the WIMP to the
nucleus and $Q_0=1.5/(M R_0^2)$. The authors of Ref.\cite{Ahlen,Freese}  determined the nuclear radius R$_0$ as:
\[ R_0 \simeq [0.91 A^{1/3} + 0.3]10^{-13} \mbox{cm} .\]

2. The form of the two parameter Fermi (2PF) distribution is
~\cite{JDL,Duda:2006uk}:
\begin{equation}\label{2PF}
\rho(r)=\frac{\rho_0}{e^{(r-c)/a}+1},
\end{equation}
where $\rho_0$ is equal to 2$\rho(r)$ at $r=c$ and
the surface thickness $t=4a\ln3$. The parameters $c$ and $a$ for
different nuclei have been determined by fitting the elastic electron scattering
experiments and muonic atom spectroscopy~\cite{data1995}. The
Fourier transform of the 2PF density Eq.~(\ref{2PF}) cannot be given as
an analytical expression, but can only be presented
numerically. In literature~\cite{Duda:2006uk} it is  named as the 2PF
form factor.

3. The `folding' charge distribution was given by
Helm~\cite{Helm} as:
\begin{equation}
\rho(r)=\int \rho_{0}(r') \rho_1(r-r') d^3 r',
\end{equation}
with certain normalization conditions.
The nuclear charge density $$\rho(r)=\rho_0=3 Ze / 4\pi R^3, \;\;\; r\leq R$$ is an approximately
uniform distribution inside a cutoff radius R, and zero outside.
The distribution $\rho_1 (r)=(1/(2 \pi s^2)^{3/2}) \exp(-r^2
/2s^2)$ is introduced to take care of the soft edge of the nucleus where $s$ is the
surface thickness. Convoluting $\rho(r)$ and $\rho_1(r)$, one deduces
the Helm form factor~\cite{Engel}:
\begin{eqnarray}
F(q)&=& \frac{3j_1(qR)}{qR}e^{-(qs)^2 /2} \nonumber\\
&=&3\frac{\sin(qR)-{qR}\cos(qR)}{(qR)^3}e^{-(qs)^2 /2},
\end{eqnarray}
with $j_1$ being the first-order spherical Bessel function. The Helm form factor
is the  form factor which is most commonly used in directly searching for dark
matter~\cite{Duda:2006uk,J.Engel,Golwala}. Lewin et al.~\cite{JDL}
proposed a set of parameters which are fixed by fitting  the muonic atom
spectroscopy~\cite{data1995} as:
\begin{eqnarray}
R^2&=&c^2+\frac{7}{3}\pi^2 a^2-5s^2\\
c  & \simeq &1.23A^{1/3}-0.6 \quad \mbox{fm}\\
s  & \simeq & 0.9 \quad \mbox{fm},\quad a\simeq 0.52 \quad \mbox{fm}.
\end{eqnarray}

4. Sick~\cite{SOGmodel} proposed to use a sum of the Gaussians (SOG)
to express the nuclear charge density:
\begin{equation}
\rho(r)=\frac{Ze}{2\pi^{3/2} \gamma^3} \sum_{i=1}^{N}
\frac{Q_i}{1+2R_{i}^2/\gamma^2}
\{e^{-(r-R_i)^2/\gamma^2}+e^{-(r+R_i)^2/\gamma^2} \},
\end{equation}
where $\gamma$ is the Gaussians width. $R_i$
and $Q_i$ are the position and charge fractions of the
i$-th$ Gaussian function with $\sum_{i} Q_i =1$. The three parameters
are determined in the elastic electron scattering
experiments~\cite{data1987}. The analytical form of the form factor is given as
~\cite{Duda:2006uk}:
\begin{equation}
F(q)=e^{-q^2 \gamma^2 /4}\sum_{i=1}^{N}
\frac{Q_i}{1+2R_i^2/\gamma^2} [\cos(qR_i)+\frac{2R_i ^2}{\gamma^2}
\frac{\sin(qR_i)}{qR_i}].
\end{equation}

5. In another model, the charge density  is described as a sum of
the Fourier-Bessel functions up to a certain cutoff radius R. The
coefficients $a_\nu$ of the Bessel expansions are directly extracted
from the elastic electron scattering experiments.  The charge
density and form factor are given~\cite{FBmodel,Duda:2006uk}:
\begin{eqnarray}
\rho(r) = \left\{
\begin{array}{ll}
\sum_{\nu=1}^N a_\nu j_0 \left(\nu \pi r / R \right) & r \le R, \\
 0& r \ge R,
\end{array}
\right.\\
F(q)=\frac{\sin(qR)}{qR}\frac{\sum_{\nu=1}^{N} (-1)^N a_{\nu}
/(\nu^2 \pi^2 -q^2 R^2)}{\sum_{\nu=1}^{N} (-1)^\nu a_\nu /(\nu^2
\pi^2)}.
\end{eqnarray}

\subsection{Nuclear density in the RMF theory }
We employ the RMF theory to calculate the nuclear density in order to
get the form factor for several nuclei which may be used as the materials of
the dark matter detector.
The RMF theory, which was first introduced by Walecka~\cite{Walecka},
has been successfully applied in various fields of nuclear
physics, especially the description of nuclear ground-state
observables~\cite{P.Ring1990,Ring:1996qi,Meng:2006,LL}.
We adopt the RMF theory with nonlinear $\sigma$ and $\omega$ terms,
which are known to provide an excellent description of the ground
states of finite nuclei including unstable nuclei~\cite{ST1994,Shen:2006nv}.
The Lagrangian for an ordinary nucleus in the RMF theory can be written
as~\cite{ST1994}:
\begin{eqnarray}
{\mathcal{L}} &=&
\bar{\Psi}_N(i\gamma^{\mu}\partial_{\mu}-M_N)\Psi_N
       +\frac{1}{2}\partial_{\mu}\sigma\partial^{\mu}\sigma-\frac{1}{2}m_{\sigma}^2\sigma^2
       -\frac{1}{3}g_2 \sigma ^3 -\frac{1}{4}g_3 \sigma^4
       -g_{\sigma }\bar{\Psi}_N\sigma\Psi_N \nonumber\\
& &
       -\frac{1}{4}\Omega_{\mu\nu}\Omega^{\mu\nu}+\frac{1}{2}m_{\omega}^2\omega_{\mu}\omega^{\mu}
       +\frac{1}{4}c_3 (\omega_{\mu}\omega^{\mu})^2
       -g_{\omega}\bar{\Psi}_N\gamma^{\mu}\Psi_N\omega_{\mu} \nonumber\\
& &
       -\frac{1}{4}\vec{R}_{\mu\nu}\vec{R}^{\mu\nu}+\frac{1}{2}m_{\rho}^2\vec{\rho}_{\mu}\cdot\vec{\rho}^{\mu}
       -g_{\rho}\bar{\Psi}_{N} \gamma_\mu \vec{\rho}^{\mu}\cdot \vec{\tau}\Psi_{N} \nonumber \\
& &
       -\frac{1}{4}F_{\mu\nu}F^{\mu\nu}
       -e\bar{\Psi}_N\gamma_{\mu}I_{c}A^{\mu}\Psi_N,
\end{eqnarray}
with
\begin{eqnarray}
\Omega_{\mu\nu} &=&
 \partial_{\mu}\omega_{\nu}-\partial_{\nu}\omega_{\mu},\\
\vec{R}_{\mu\nu} &=&
 \partial_{\mu}\vec{\rho}_{\nu}-\partial_{\nu}\vec{\rho}_{\mu}, \\
F_{\mu\nu} &=& \partial_{\mu}A_{\nu}-\partial_{\nu}A_{\mu},
\end{eqnarray}
where $\Psi_N$ is the nucleon field with the mass $M_N$, $\sigma$,
$\omega_{\mu}$, $\vec{\rho}_{\mu}$ are defined as meson fields with
masses of $m_\sigma, m_\omega, m_\rho$. $g_{\sigma} $, $g_{\omega} $
and $g_{\rho} $ are respectively the couplings $\sigma NN$, $\omega
NN$, and $\rho NN$. $g_2$, $g_3$ and $c_3$ are coupling constants of
the self-coupling terms for the $\sigma$ and $\omega$ mesons.
$A_\mu$ is the electromagnetic (EM) field. $\vec{\tau}$ is the
nucleon isospin operator and $I_{c}=(1+\tau_{3})/2$ is the projector
and $\tau_3$ is the third component of the isospin Pauli matrices
for nucleons. In the RMF theory, $\sigma$, $\omega$, $\rho$ and the
EM field $A$ are classical and treated as  mean fields. In the
Lagrangian, there are five independent fields: $\Psi_N$, $\sigma$,
$\omega$, $\rho$, and the EM field $A$. Employing the Euler-Lagrange
equation, one can obtain five coupled equations corresponding to the
five independent fields. Directly solving the equations is
impossible because they are non-linear, thus one needs to introduce
suitable approximations. These coupled equations can be solved in a
standard way within the RMF approach~\cite{P.Ring1990}.

We deal with static and spherically symmetric nuclei, and the meson
fields depend only on radius, then the zero components $\omega^0$,
$\rho^{03}$ and $A^0$ would make the dominant contributions. The
equation of motion for the nucleon field is the Dirac equation where
the classical fields $\sigma$, $\omega$, $\rho$, and $A$ stand as
the effective potential for $\Psi_N$,
\begin{equation}\label{Dirac}
\{i\gamma_{\mu} \partial^{\mu}-[M_N+g_{\sigma} \sigma(r)
]-g_{\omega}\gamma_{0} \omega^0(r)-g_{\rho} \gamma_{0} \tau_3
\rho^{03}(r) - e\gamma_0 I_c A^0(r)\}\Psi_{N}({\bf r},t)=0.
\end{equation}
In the RMF theory the baryon field is still quantum. One can input a
trial potential into Eq. (\ref{Dirac}) and solve  $\Psi_N$, then
substitute this solution into the equations for the meson and EM
fields as the source terms to solve those fields. Iterating many
times, the solutions would converge. The trial potential in
principle can be freely chosen, but the convergence rate depends on
the form of the trial potential. One can always choose a potential
according to his knowledge on the nuclear structure or a reasonable
intuition.

We seek for solutions for $\Psi_{N}({\bf r},t)$ which have
fixed energies. The nucleon field operator $\Psi_N({\bf r}, t)$ can be
written as a sum of positive and negative solutions $u_\alpha({\bf
r})$ and $\nu_\alpha({\bf r})$ where $\alpha$ stands for the
complete quantum numbers of an energy eigenstates and in the second
quantization picture the wave function of a fermion can be written as
\begin{equation}
\Psi_N({\bf r},t)=\sum_{\alpha}[a_{\alpha}u_{\alpha}({\bf r})e^{-i
E_{\alpha} t}+b_{\alpha}^{\dag}\nu_{\alpha}({\bf r})e^{i
E_{\alpha}t}].
\end{equation}
$a_{\alpha}$ and $b_{\alpha}^{+}$ are annihilation and creation
operators of baryon and anti-baryon. The subscript $\alpha$ runs
over the full set of single-particle quantum numbers. For the normal
nuclei, generally the negative solution $\nu_{\alpha}({\bf r})$ can
be ignored in the $no~sea~approximation~\cite{P.Ring1990}$ . The
probability operator of the nucleon density at ${\bf r}$ is
\begin{equation}\label{nucleon}
\Psi_N^{\dagger}({\bf r},t)\Psi_N({\bf r},t).
\end{equation}

The nuclear ground state can be expressed as
$$|\Psi_{A}>=(\Pi_{i=1}^A a_i ^{\dag}  )|0>,$$
where $|0>$ is the vacuum, A is the number of nucleons in the
nucleus and the product corresponds to creating A nucleons (protons
and neutrons) which are the energy eigenstates. Sandwiching  the
probability operator Eq. (\ref{nucleon}) between the nuclear ground state,
we obtain the nuclear density as
\begin{equation}
\rho_{B}({\bf r})=\sum_{\alpha}u_{\alpha}^{\dag}({\bf r})
u_{\alpha}({\bf r}).
\end{equation}

\begin{figure}
\begin{center}
  \includegraphics[width=12cm]{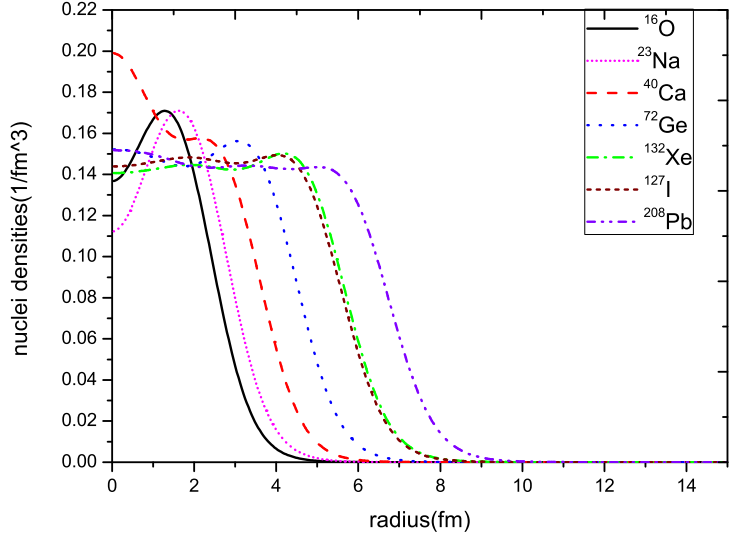}\\
  \caption{Nuclear densities of $^{16}$O,  $^{23}$Na, $^{40}$Ca,
  $^{72}$Ge, $^{127}$I, $^{132}$Xe, $^{208}$Pb, obtained in terms of the RMF theory.
}\label{1}
\end{center}
\end{figure}

The input parameters are $M_N=938$ MeV, $m_{\sigma}=511.198$ MeV,
$m_{\omega}=$783 MeV, $m_{\rho}=770$ MeV, $g_{\sigma}=10.0289$,
$g_{\omega}=12.6139$, $g_{\rho}=4.6322$, $g_2 =-7.2325$,
$g_3=0.6183$, and $c_3=71.3075$, called as the TM1 parameter set,
which was determined by fitting some experimental data of finite
nuclei in Ref.~\cite{ST1994}. This parameter set is able to describe
many ground state properties of finite nuclei all over the periodic
table. We calculate the nuclear densities of several nuclei from
light to heavy. In Fig. (\ref{1}), we show the results of nuclear
densities of $^{16}$O,  $^{23}$Na, $^{40}$Ca, $^{72}$Ge, $^{127}$I,
$^{132}$Xe, $^{208}$Pb \footnote{It is well known that the RMF
theory is not a
  good working framework for even-odd nuclei, such as $^{23}$Na and  $^{127}$I, but since those elements are possibly chosen as
  the dark matter detector materials, we use the RMF approach to evaluate their form factors as a reference for further studies. We will investigate
  their form factors in a more proper framework, if such elements are indeed chosen to be the detector materials. }.

Our next task is to perform  the Fourier
transformation of the RMF density numerically to gain the form factor.

The kinematics for the collisions between WIMP and nucleus: the recoil energy is
\begin{equation}
E={1\over 2}mv^2{2mM_A\over
(m+M_A)^2}(1-\cos\theta_{cm})={m_{r}^2v^2\over
M_A}(1-\cos\theta_{cm}),
\end{equation}
and the recoil three-momentum is
\begin{equation}
|\vec q|=\sqrt{2M_A E},
\end{equation}
where $m$ and $M_A$ are the masses of the WIMP and nucleus
respectively, $m_{r}$ is the reduced mass, $v$ is the velocity of
the WIMP and $\theta_{cm}$ is the scattering angle in the center of
mass frame~\cite{ChenSL}. In the lab frame, the recoil momentum is
exactly the momentum transferred from the incident WIMP to the
nucleus which is supposed to be at rest before impact of the WIMP.

\begin{figure}
\begin{center}
  \includegraphics[width=12cm]{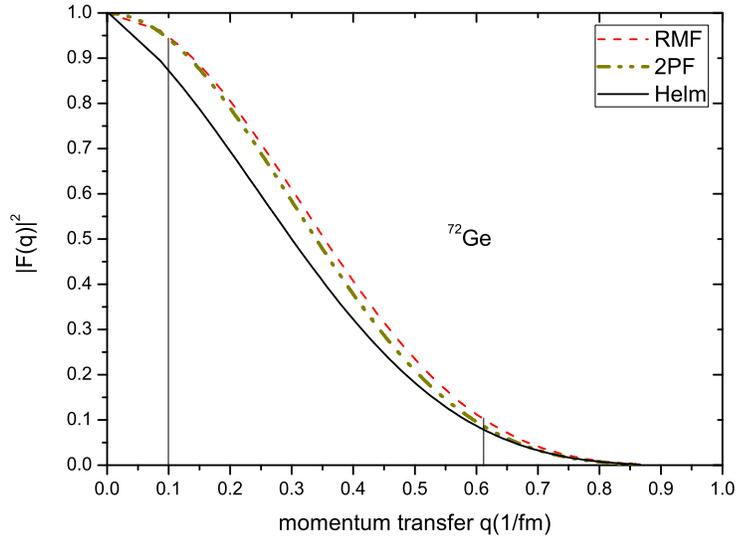}\\
  \caption{RMF, 2PF and Helm form factors for $^{72}$Ge for the recoil energy range from 10 keV (q $\simeq$ 0.19 fm$^{-1}$ ) to 100 keV (q $\simeq$ 0.61 fm$^{-1}$ ).}\label{2}
\end{center}
\end{figure}

Taking Ge as an example, in Fig. (\ref{2}) we show the dependence of
the RMF form factor on the recoil momentum. For a comparison we also
draw the curves corresponding to the Helm and 2PF form factors. The
relevant parameters of the Helm and 2PF are taken from
Refs.~\cite{Duda:2006uk,data1995}. As can be seen from the Fig.
(\ref{2}), the results of RMF  are 2\% to 8\% above that of 2PF form factors at
most of the energy region from 2 keV (momentum q $\simeq$ 0.1 fm$^{-1}$) to
200 keV (q $\simeq$ 0.9 fm$^{-1}$), and lower by less than 1\%
at the range of $q\leq 0.1$ fm$^{-1}$. But at all the recoil energy regions, the RMF results
are above the Helm form factor by about 15\% to 25\%. At 1 fm$^{-1}$, all of them attenuate to zero, or very close to zero.

\begin{figure}
  \includegraphics[width=9cm]{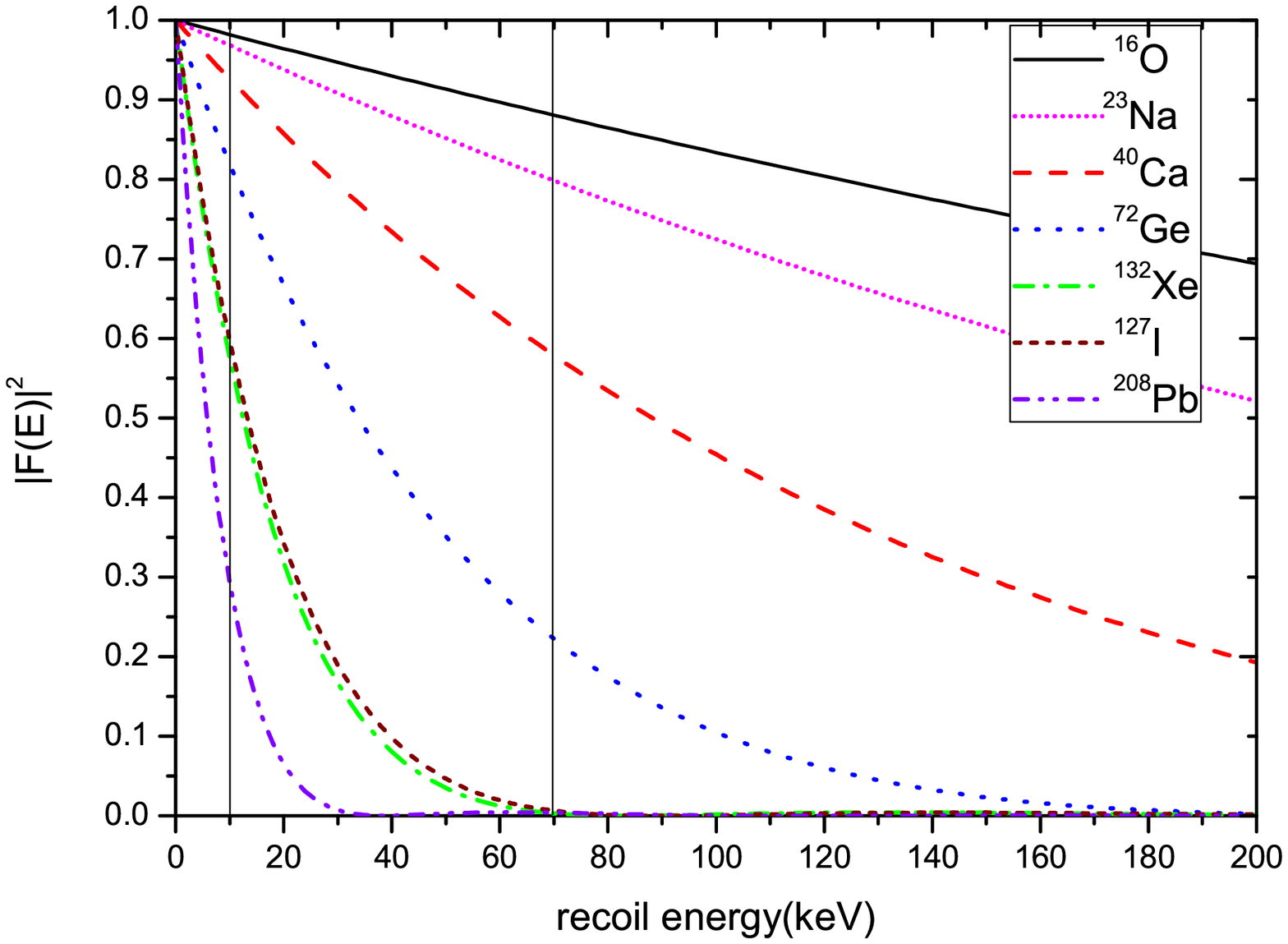}
  \includegraphics[width=9cm]{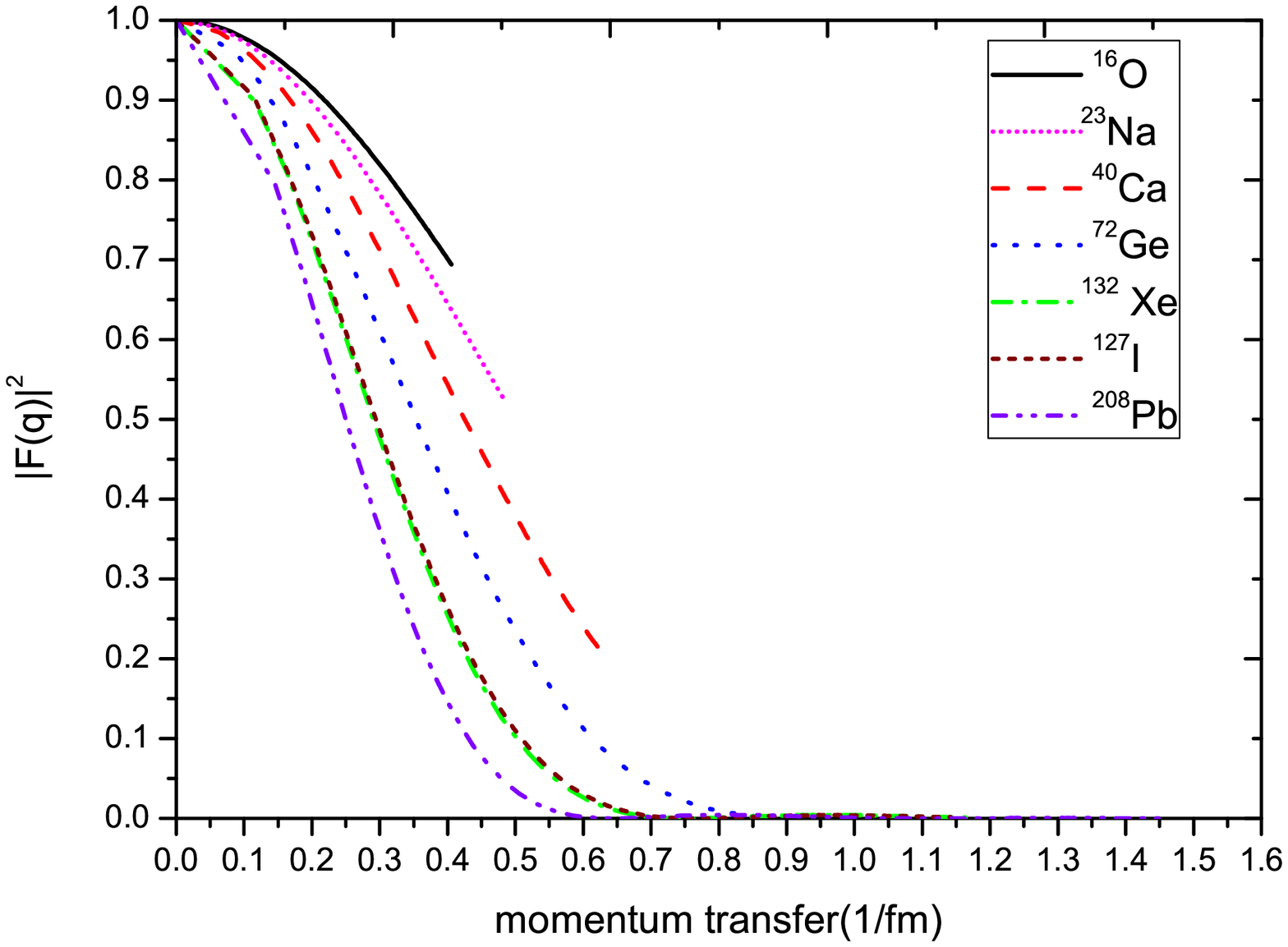}
  \caption{The RMF form factors for various nuclei corresponding to the nuclear densities shown in figure 1.
The left and right pictures are the dependence of the factors as functions on respectively the recoil energy and the recoil momentum.}\label{3}
\end{figure}

We plot the dependence of the RMF form  factors on the recoil energy
and recoil momentum (for a clearer illustration) for various nuclei in Fig. (\ref{3}).
It is shown that the form factor generally drops faster as the
nucleus mass increases at higher recoil energies.  Among the
commonly adopted detector materials, F$_{Xe}$(E)(dash dot line) and
F$_{I}$(E)(short dash line) have similar behaviors. They drop off
more rapidly at the recoil energy range from 10 keV to 100 keV and
F$_{Xe}$(E) tends to zero at 60-70 keV. It significantly influences
the detection sensitivity of the Xenon detector. At the recoil energy larger
than 100 keV, F$_{I}$(E) and F$_{Ge}$(E)(dot line) are too small to
perform effective detections. In the case of Na and Ca, the form
factors decrease more slowly in the whole energy spectrum.

\section{Discussion and Conclusions }
Though the RMF form factor is not analytical, its advantage is
obvious. We are based on the theory which has been proved to be
successful for dealing with nuclear problems, especially determining
the nuclear densities for spherically or near-spherically symmetric
nuclei. In the RMF theory, by resolving the Dirac equation to get
nucleon  wave function and with it the nuclear density can be
obtained. Since this method does not depend on the initial form of
the potential, even though we need to choose one for the first
input, it has obvious advantage. The iteration would smear the
difference between different inputs and make the final solution to
converge to a should-be reliable one. Therefore we believe that the
RMF form factor would be a more accurate form factor for the
calculation of dark matter detection, and moreover, it also
determines the energy-sensitivity curve of the material adopted for
the detector.

The results can be utilized as a reference for designing dark matter
detectors and make accurate extraction of information about the
identity of the WIMP and its interaction with nucleon, or even with the
ingredients inside nucleons (say, quarks) from
data~\cite{QiaoQP,HeXG,LiTong,Fengtf,Feng:2006ni}.

For readers' convenience, besides the figures we presented above, we
make a table where some numbers are explicitly listed. As can be
seen in Table 1, at the recoil energy ranges, the form factors of the three models
have ceratin differences, the RMF and 2PF form factors
differ from each other by only 4\%-8\%, but differ by at least
15\% to 25\% from the Helm form factor. The Fig. (\ref{3}) tells us from 10
keV to 70 keV, $F_{Xe}(E)$ obviously declines from $F_{Ge}(E)$ and
$F_{I}(E)$.
\begin{table}[h]
\caption{the three form factors at 10 keV, 50 keV, 100 keV
, for different nucleus.}
\begin{center}
\begin{tabular}{|c|c|c|c|c|c|c|c|c|}
      \hline
      $E    $  & $|F(E)|^2$  & O16   & Na23      & Ca40   & Ge72  & I127   & Xe132  & Pb208 \\
      \hline
      10 keV   & RMF         & 0.982 & 0.968     &0.926   &0.818  & 0.596 &0.576  &0.293\\
      \cline{2-9}
               & 2pF         &0.981   & 0.965    &0.920   &0.802  & 0.605 &0.579  &0.305\\
      \cline{2-9}
               & Helm        &0.908   & 0.882    &0.825   &0.710  & 0.507 &0.489  &0.251\\
      \hline
      50 keV   & RMF         &0.913   & 0.852    &0.678   &0.351  & 0.047 &0.035  &0\\
      \cline{2-9}
               & 2pF         &0.907   & 0.838    &0.658   &0.324  & 0.052 &0.042  &0\\
      \cline{2-9}
               & Helm        &0.777   & 0.706    &0.545   &0.278  & 0.040 &0.031  &0\\
      \hline
      100 keV  & RMF         &0.834   & 0.704    &0.454   &0.105  & 0     &0       &0\\
      \cline{2-9}
               & 2pF         &0.809   & 0.700   &0.429   &0.098   & 0     &0       &0\\
      \cline{2-9}
               & Helm        &0.671   & 0.567   &0.346    &0.083   & 0    &0       &0\\
      \hline
\end{tabular}\label{4}
\end{center}
\end{table}

Here, We assume the target nuclei discussed in this work to be spherical or
near-spherical, however realistic nuclei may be slightly deformed.
Such deformations
possibly  affect the shape of the form factors and we are going to discuss the effects in our next work.\\

\noindent{Acknowledgements}:

This work is supported by the National Natural Science Foundation of China.

\end{document}